# Early detection of knee osteoarthritis using deep learning on knee magnetic resonance images

Anastasis Alexopoulos[1], Jukka Hirvasniemi[2,*], Nazlı Tümer[1]

[1]Department of Biomechanical Engineering, Delft University of Technology (TU Delft), Delft, The Netherlands;
[2]Department of Radiology and Nuclear Medicine, Erasmus MC University Medical Center, Rotterdam, The Netherlands.

*corresponding author: j.hirvasniemi@erasmusmc.nl

**Abstract.** An increasing number of deep learning techniques have been used to detect osteoarthritis (OA) and study its incidence and progression. Although valuable imaging-based biomarkers for OA are typically derived from magnetic resonance imaging (MRI), most of the deep learning techniques have been developed based on plain radiography (X-ray) images of patients and their demographics (*e.g.*, age, gender, body mass index (BMI)). In this study, we aim to investigate the influence of MRI and patient data on the prediction of knee OA incidence using different deep learning architectures. Knee OA incidence within 24 months was predicted using the intermediate-weighted turbo spin-echo (IW-TSE) sequence of 593 patients from the Osteoarthritis Initiative. To extract a region of interest containing the knee joint from the IW-TSE sequence, a U-Net model was trained and used to segment bone on a dual echo steady state (DESS) sequence. Subsequently, IW-TSE and DESS sequences were registered and the DESS segmentations were transformed to the corresponding IW-TSE scans. The performance of MRI-based features in the prediction of knee OA incidence was tested using three different deep learning architectures: a residual network (ResNet), a densely connected convolutional network (DenseNet), and a convolutional variational autoencoder (CVAE). To evaluate the predictive performance of MRI-based features alone, the outputs of ResNet, DenseNet, and CVAE were coupled with patient data (*i.e.*, age, gender, BMI) and used as input to a Logistic Regression (LR) Classifier. Knee OA was defined based on visual MRI and X-ray-based OA features. The performance of the segmentation was evaluated using the Dice similarity coefficient, while those of the OA detection algorithms were assessed using the area under the receiver operating characteristic curve (AUC) and the precision-recall curve (PR-AUC) metrics. Regarding the segmentation of tibial and femoral bones, the Dice similarity coefficients were 0.985 and 0.987, respectively. The ResNet and DenseNet showed similar results, with both methods having AUC values up to 0.6269. The best performing OA detection model was CVAE with an AUC of 0.6699 when combined with patient data and an AUC of 0.6689 when used alone as input to the LR classifier. All three detection algorithms yielded higher performance metrics when patient data were combined with MRI-based features. The results showed that three deep learning algorithms have similar metrics when using IW-TSE MRIs and their performance increased with the inclusion of patient data, which shows the strong influence of variables such as age, gender, and BMI on the detection of knee OA.

**Introduction**

Osteoarthritis (OA), which affects more than 250 million people annually worldwide [1], is characterized by the degeneration of articular cartilage and bone where the intrinsic repair mechanisms are insufficient. While any synovial joints can be affected by OA, the knee joint is one of the most affected sites of OA development [2], [3]. Several structures of the knee joint are affected by OA, leading to articular cartilage loss, bone remodelling, synovitis, and lesion development in the bone marrow [4]. The structural changes in the joint are the basis of OA detection and diagnosis, along with the presence of symptoms such as joint stiffness and pain, even though the structural changes are detected when the disease is advanced and irreversible [4], [5]. The early identification of cartilage loss and/or subchondral bone degeneration in patients without or with a few symptoms could help clinicians plan effective treatment strategies [5].

To depict the joint structures and structural changes, several imaging modalities can be used. One of the imaging methods used is plain radiography (X-ray). Although X-ray images can be acquired in a few seconds and can allow the extraction of morphological and statistical features, they hinder the detection of early developed OA-related changes, such as localized cartilage degradation, because cartilage tissue is not visible on X-ray images and they are 2D projections of 3D structures [5], [6]. Another imaging modality is magnetic resonance imaging (MRI) with the ability to create 3D visualizations of the knee joint (*e.g.*, subchondral bone, cartilage, ligaments), which makes the MRI scans more suitable for the early detection of knee OA when compared to X-rays.

Recent advancements in the field of artificial intelligence and deep learning have led to the incorporation of algorithms in the analysis of medical images and patient data. Several deep learning algorithms have been developed to detect knee OA and study its severity, incidence, and progression. Many of the previous deep learning studies for OA assessment have used X-ray images and focused on the classification of knee joints based on their OA severity [7]. Few studies have used deep learning to predict the incidence and progression of knee OA from X-rays [8]. MRI has not been yet investigated thoroughly for its impact on the early detection of OA, even though this imaging modality can enable the detection of bone and cartilage changes that may increase the risk of joint collapse at short periods of time (1-2 years) [9], [10]. One study used deep learning and MRI data to predict X-ray based joint space narrowing within 12 months [11]. However, MRI-based outcome could provide a more comprehensive view of OA. Therefore, the aim of this study was to investigate the ability of deep learning algorithms to predict knee OA incidence within 24 months using MRI and demographic variables.

**1. Materials and Methods**

The steps followed to develop deep learning methods for the early prediction of OA are explained in detail in this section. The main steps include *(1)* MRI and patient data collection, *(2)* post-processing of MRI scans, and *(3)* development of deep learning algorithms for the early detection of OA.

**1.1. Data collection**

Two datasets, *i.e.*, *(1)* knee MRI (kMRI) semi-quantitative (SQ) MOAKS and *(2)* kMRI SQ WORMS, that are publicly available in the database of Osteoarthritis Initiative (OAI) were examined. The first dataset contains MRI scans of patients at five different time points (*i.e.*, baseline, 12-, 24-, 36-, and 48-month visits), while the second dataset consists of MRIs of patients at three different time points (*i.e.*, baseline, 24-, and 48- month visits). The structural abnormalities in the knee joint were assessed by using the MOAKS and WORMS grading systems in the first and second datasets, respectively. According to the MOAKS grading scale [12], tibiofemoral OA has been defined based on the presence of both group [A] features or one group [A] and two or more group [B] features: *i.e.*, Group A – definite osteophyte formation >= 2, full thickness cartilage loss >= 3; Group B – subchondral bone marrow lesion or cyst >= 1, meniscal subluxation, maceration or degenerative tear >= 2, partial thickness cartilage loss (where full thickness loss is not present) >= 1. Besides, according to WORMS [13], OA in the whole knee joint is diagnosed if the following items are fulfilled: *i.e.*, cartilage morphology score >= 3, bone marrow lesions (BML) score >= 2, osteophytes >= 2. Considering these gradings, both datasets were screened to find subjects who had the features reported and developed OA between specific visits. As neither of the datasets had MRI-based osteophyte scores available, X-ray-based osteophyte grades were used. This approach was taken, as osteophytes depicted in X-rays have been reported to correlate with cartilage damage in MRIs [14]–[16]. Since the MOAKS grading is newer and refined version of the WORMS grading, MRIs of patients with and without OA (*i.e.*, control) at two follow-up visits, *i.e.*, 12- and 24- months, were retrieved from both datasets. Along with the sagittal three-dimensional dual-echo in a steady state with water excitation (SAG 3D DESS WE) and the coronal intermediate weight two-dimensional turbo spin-echo (COR 2D IW-TSE) MRI sequences, patient data including age, gender, and BMI at baseline time point were retrieved from the OAI database.

**1.2. Postprocessing of MRI scans**

The area around the knee joint was semi-automatically extracted from each IW-TSE MRI scan. The extraction process of the region of interest (ROI) consisted of two steps: *(1)* segmentation, *i.e.*, the identification of pixels that belong to the tibia and femur, and *(2)* the definition of the area surrounding the knee joint. As annotated 3D DESS MRI scans had been already available to the authors, the DESS sequence was used to segment the knee joint. Besides, the IW-TSE MRI sequence was used for the early detection of knee OA due to its sensitivity and specificity to bone and cartilage changes.

*1.2.1. Segmentation of DESS MRI scans*

The segmentation of DESS MRI scans was achieved with the application of the U-Net method [17]. The U-Net model consisted of four down-sampling and up-sampling steps to extract spatial and context information, *i.e.,* where each pixel is and to which label, tibia or femur, this pixel belongs. Each down-sampling step was composed of two 2D convolution layers, one dropout layer and one max-pooling layer, while each up-sampling step included one transpose convolution, one dropout layer, and two convolution layers (Figure 1). In the training of the U-Net model, the categorical cross-entropy loss function together with Adam's optimizer was used. The performance of the U-Net model in segmenting tibia and femur bones was assessed using the Dice similarity coefficient and Intersection of Union (IoU).

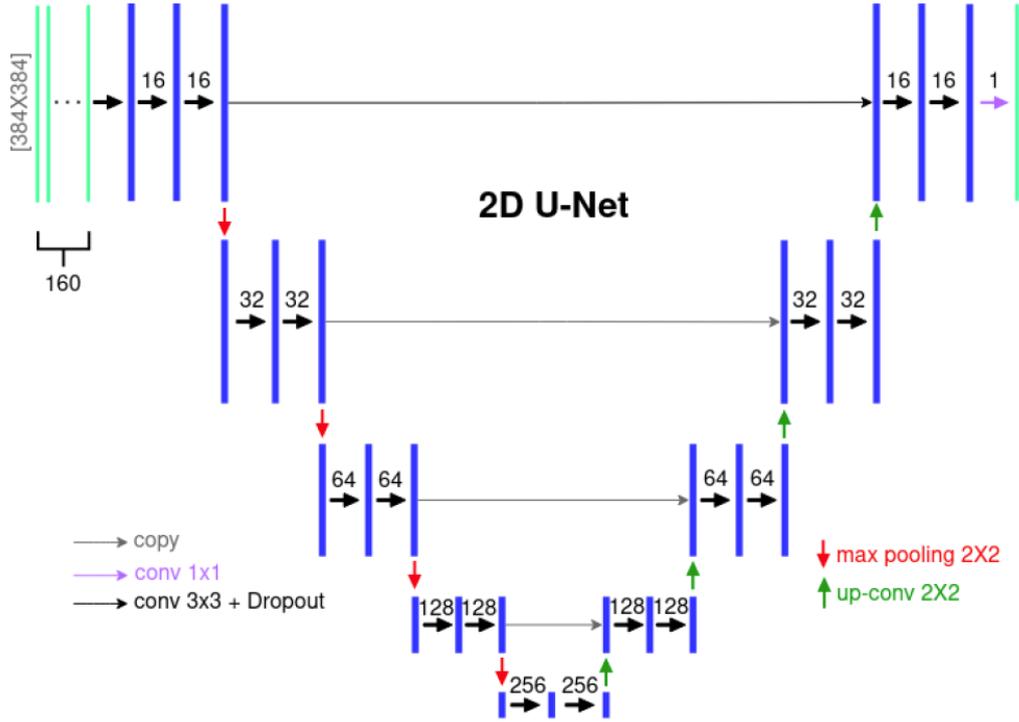

**Figure 1.** A 2D U-Net model for the segmentation of the sagittal 2D DESS MRI scans.

To train and validate the U-Net model, segmentations of 3D DESS MRI scans of 507 patients were retrieved from the publicly available OAI-ZIB dataset [18]. The acquisition parameters of these DESS MRI scans were the following: matrix phase: 307, matrix frequency: 384, the field of view: 140 mm, slice thickness/gap: 0.7 mm / 0 mm, and flip angle: 25°. Following the training and validation of the U-Net model, the DESS MRI scans retrieved from the OAI database were given as input to the model to create segmentation masks (*i.e.*, each pixel was classified into five classes: *(1)* background, *(2)* the femur, *(3)* the femoral cartilage, *(4)* the tibia, and *(5)* the tibial cartilage.

### 1.2.2. Definition of the area around the knee joint

With the aim of creating the IW-TSE segmentation masks and extracting the area around the knee joint, DESS MRIs were registered to IW-TSE MRIs using Elastix [19]. The following steps were repeated for each patient involved in this study: *(1)* DESS MRI scan of a patient with a matrix size of 384 x 384 and 160 slices was registered through an affine transformation to the IW-TSE MRI scan of the same patient with a matrix size of 384 x 384 and 37 slices, and *(2)* the transformation parameter map that was obtained following the registration was applied to the DESS segmentation masks to describe those in the IW-TSE MRI scan. Each registration result, *i.e.*, the superimposition of an IW-TSE MRI scan with its segmentation masks, was visually checked to ensure its correctness. During the visual checks, it was observed that several pixels were misclassified in some MRIs, *i.e.*, background pixels classified as bone pixels. To discard these misclassified pixels, the Canny edge detection algorithm [20] was applied to the IW-TSE MRI segmentation masks and pixels that belonged to bones were separated from those of the background. The left side IW-TSE MRIs were mirrored as they were right side, and the minimum and maximum coordinates of bone pixels across the IW-TSE MRIs were determined. Following this step, the coronal slices that do not depict soft tissues or image background alone but do cover bone tissues (*i.e.*, more than 500 pixels describing the bone tissue) were selected. This step resulted in the reduction of coronal plane slices in the initial 3D IW-TSE MRIs. Instead of the initial 37 coronal slices, the new images contained coronal slices ranging from 18 to 27. To standardize the shape of the image inputs for the early OA detection algorithms, 3D images with coronal slices greater than 18 were interpolated using the SciPy Python image processing library. The final MRI images were cropped to extract the area around the knee joint. The final IW-TSE MRI scans had 250 x 320 pixels and 18 slices.

## 1.3. Deep learning algorithms for early detection of OA

In this study, Residual Network (ResNet) [21], Densely Connected Convolutional Network (DenseNet) [22], and Convolutional Variational Autoencoder (CVAE) [23] architectures were developed and evaluated for their ability to detect the presence of OA. In the following subsections, the implementation details of these mentioned deep learning algorithms are presented.

### 1.3.1. ResNet

Several variations of ResNet have been developed, the shallower versions (*i.e.*, ResNet-18, ResNet-34) with two 3 x 3 convolutional layers in the residual blocks (Figure 2a) and the deeper versions (*i.e.*, ResNet-50, ResNet-101, ResNet-152) with two 1 x 1 and one 3 x 3 convolutional layers (Figure 2b). For practical reasons (*i.e.*, computational resources), the ResNet-50 was chosen. ResNet-50 (Figure 2c) is composed of 50 layers and of four different residual blocks, with each block repeated several times (*i.e.*, 3, 4, 6, 3, respectively). Each repetition of the residual blocks is applied with a different number of filters, and it is concluded by compressing the output feature map to match the input dimensions of the next repetition. The implemented ResNet-50 was modified to be able to receive the desired 3D IW-TSE MRI inputs. The total number of learned parameters was around 45 million and was trained with the use of the binary cross-entropy function, which gives as an output the probability of a single 3D image belonging to the OA progress group.

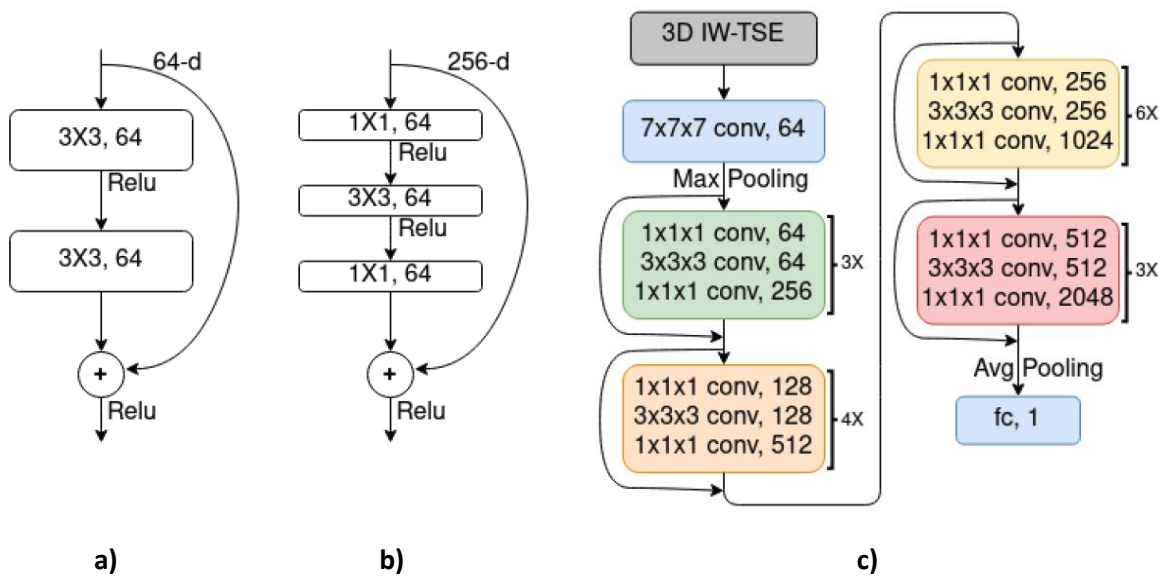

**Figure 2. a)** Residual blocks for ResNet-18 and ResNet-34; ReLU: Rectified linear unit **b)** Residual blocks for ResNet-50, ResNet-101 and ResNet-152; ReLU: Rectified linear unit, **c)** The ResNet-50 architecture used in this study; conv: convolution, avg: average.

*1.3.2. DenseNet*

DenseNet constitutes a more elaborate and complicated version of ResNet. DenseNet utilizes the outputs of several convolutional layers through feature reuse, thus resulting in an easier-to-train and parameter-efficient network. The main difference between DenseNet and ResNet is the concatenation of feature maps learned by all previous layers, which increases the variation in the input values of subsequent layers. The building operation of the DenseNet is called composite function and includes three layers, a batch normalization, followed by a rectified linear unit (ReLU) and a convolutional layer. Two composite functions, with 1 x 1 and 3 x 3 convolution kernel sizes respectively, construct a dense block. Four dense blocks are used to create the DenseNet architecture with specific filters and repetitions, which lead to DenseNets with different depths, *i.e.*, 121, 169, 201, and 264 layers. Between each of these blocks, a transition layer is implemented, which facilitates the down-sampling of feature maps and thus, the network's parameter efficiency. In this study, the shallower version of DenseNet, *i.e.*, DenseNet-121, was tested. This model (Figure 3) is composed of 121 layers and of four dense blocks with different feature maps and repetitions (*i.e.*, 6, 12, 24, 16) for each block. The parameters of the convolution and pooling layers were modified to be able to receive the desired 3D IW-TSE MRI inputs, resulting in approximately 11 million trainable parameters. Like the ResNet-50 model, DenseNet-121 was also trained with the incorporation of the binary cross-entropy loss function.

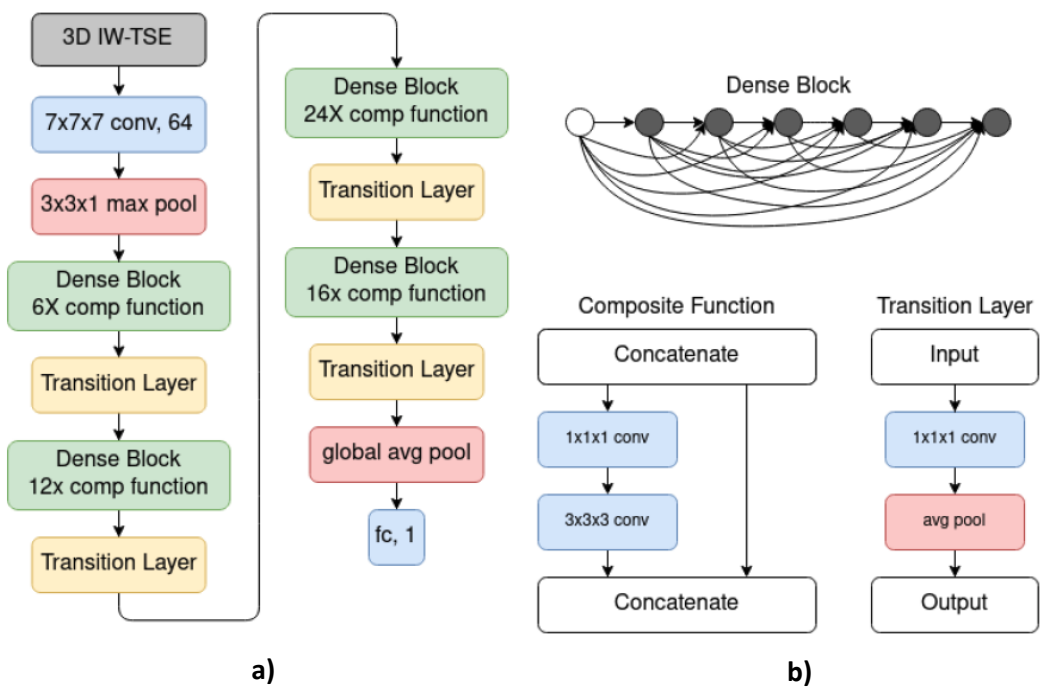

**Figure 3. a)** The 3D DenseNet-121 model, **b)** First dense block with 6 repetitions of composite function (upper), composite function with two convolution layers, transition layer with one convolution and one average pooling layer (lower); avg: average.

*1.3.3. CVAE*

Convolutional Variational Autoencoder (CVAE) is constructed with the use of 3D convolutional and de-convolutional layers for the encoding and decoding respectively, and the addition of latent space reparametrization. This latter enables the feature vector to be determined by a multivariable Gaussian distribution:

$$z = \mu_x + \sigma \times \epsilon \tag{1}$$

where $\epsilon \sim N(0, I)$, with $\mu$ being the mean, $\sigma$ the standard deviation and $\epsilon$ Gaussian Noise variable. Therefore, in the variational autoencoder, the encoder outputs a probability distribution in the bottleneck layer, from which the latent space is sampled.

Due to the complex nature of knee MRI scans and the high similarity between the range of interests in healthy and early OA progression subjects, a discriminative term [24] was incorporated into the objective function. This term, namely discriminative penalty, forces the network to extract features that minimize intra-class distances and maximize those of inter-class, and is expressed as:

$$\Omega_{disc} = \frac{\sigma_1^2 + \sigma_2^2}{|\mu_1 - \mu_2|^2} \tag{2}$$

where, $\mu_i$ and $\sigma_i^2$ are the mean and variance of the learned latent space respectively of each class. Thus, the overall loss function was:

$$J_{CVAE} = J_{MSE} + \lambda \Omega_{disc} \tag{3}$$

where, $J_{MSE}$ and $\lambda$ are a loss function of the mean square error and the discriminative penalty weight, respectively. The CVAE model (Figure 4) is constructed with three convolution and three deconvolution layers, along with two dense layers for the creation of the latent space, resulting in around 22 million parameters.

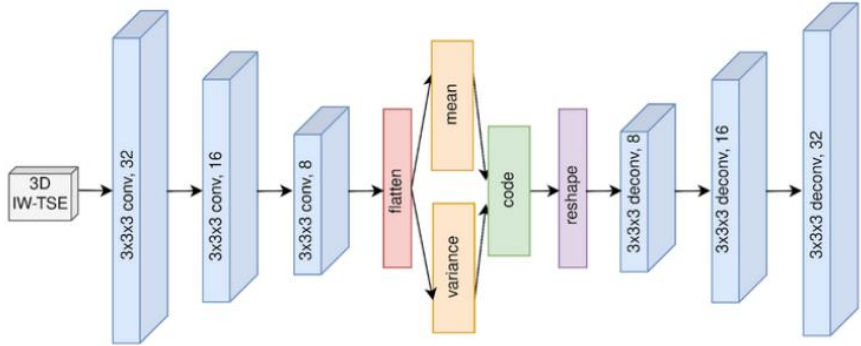

**Figure 4.** The 3D Convolutional Variational Autoencoder model; conv: convolution, deconv: deconvolution.

### 1.3.4. *Training of the deep learning models and their performance evaluation*

Several variations of the three models were obtained to counter class imbalance (*i.e.*, the number of subjects without OA vs. the number of subjects with OA) and investigate the influence of convolutional layer parameters on the models' performance. Both ResNet and DenseNet models were trained using *(1)* only original 3D IW-TSE MRIs and *(2)* a combination of original and augmented 3D IW-TSE MRIs. The augmented images were obtained by rotating (±5°) the original images and changing their contrast (x 1.5) and brightness (+50 in each pixel value). To enhance the attention of ResNet-50 and DenseNet-121 models on the under-represented OA progressing group, class weights were included in the training of original 3D IW-TSE MRIs. These class weights were calculated using an embedded function in the Scikit Learn Python library (*i.e.*, *sklearn.class_weight*). Furthermore, the variants of

ResNet-50 and DenseNet-121 models were trained with and without having an $L_2$ norm regularization term in each convolution layer, resulting in a total of five variations: ResNet-50 and DenseNet-121 models trained on *(1)* original 3D IW-TSE MRIs, *(2)* original 3D IW-TSE MRIs and $L_2$ norm regularization term, *(3)* original 3D IW-TSE, $L_2$ norm regularization term and class weights, *(4)* augmented 3D IW-TSE MRIs, and *(5)* augmented 3D IW-TSE MRIs and $L_2$ norm regularization term. In the case of the CVAE model, the only augmentation method was the oversampling (*i.e.*, creating duplicates) of the OA progression group due to the structure of the model that required the presence of the same number of subjects from each class, *i.e.*, control and OA progression groups. The combination of this with and without having the $L_2$ norm regularization term was also considered, resulting in a total of two variations: (1) CVAE model trained on *(1)* augmented 3D IW-TSE MRIs, and *(2)* augmented 3D IW-TSE MRIs and $L_2$ norm regularization term.

Patient data including age, gender, and BMI were incorporated by using a Logistic Regression classifier (LRC). While the output probability values of the ResNet-50 and DenseNet-121 models were combined in a vector with patient data, the learned feature vector of CVAE was concatenated with patient data. The training set was composed of 70% of the available 3D IW-TSE MRIs. The rest of the MRIs was equally split into two to validate (15%) and test (15%) the models. The data augmentation was applied only in the training dataset. During the training process, Adam's optimizer was used for each of the three deep learning architectures. Several different values of model hyperparameters were tested: input batch size and learning rate in both ResNet-50 and DenseNet-121 models, and epoch number, discriminative penalty term, latent space size, batch size, and learning rate in CVAE. The performance of each model along with the LRC on the testing dataset was assessed using two metrics, *i.e.*, the area under the receiver operating characteristic curve (AUC) and the precision-recall curve (PR-AUC).

## 2. Results

### 2.1. Data and its post-processing

The final dataset included the DESS and IW-TSE MRIs of 593 subjects. 73% of the subjects ($n_{control}$ = 434) did not progress in knee OA after 24 months, while the rest ($n_{OA}$ = 159) did. The U-Net model that was used for the segmentation of DESS MRIs achieved highest performance metrics with a batch size of 16. While the mean Dice coefficient was 0.99542 of all classes, the mean IoU exceeded 0.90. Therefore, the batch size of 16 was chosen to train the U-Net model and segment the 593 DESS MRIs. Following the segmentation, the Dice coefficient and the mean IoU were higher than 0.98 and 0.96, respectively, for both the femur and the tibia. Figure 5 shows two randomly selected segmentation results.

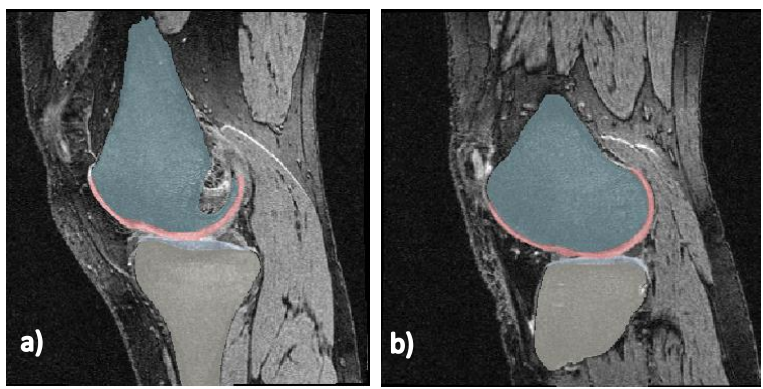

**Figure 5.** Segmentation labels (*i.e.*, dark blue mask: the femur, pink mask: the femoral cartilage, light blue: the tibial cartilage, and gray mask: the tibia) are superimposed on the corresponding DESS MRI scans. **a)** The sagittal view of the right knee of a patient within the control group, **b)** The sagittal view of the left knee of a patient with the OA progression group.

Two randomly picked examples of segmentation labels on the corresponding IW-TSE MRIs, which were obtained following the registration process, are shown in Figure 6a. In several cases (Figure 6b), some background areas were misclassified and labelled as bone regions. Going through the misclassified samples, the Canny Edge Detection parameters were determined. Accordingly, variance ($\sigma$) was 8, while the low and high bone thresholds were 0.9 and 0.95 for the tibia, and 0.37 and 0.39 for the femur.

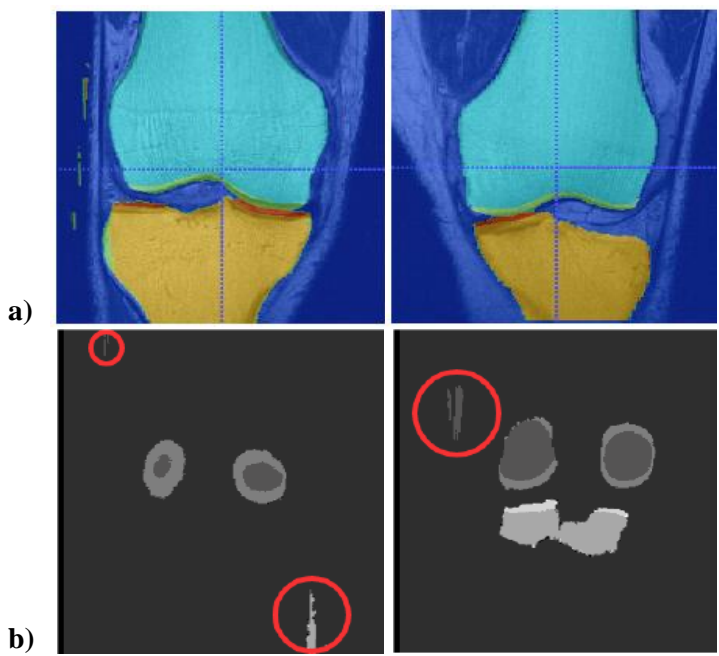

**Figure 6. a)** Alignment of IW-TSE MRIs and their corresponding IW-TSE segmentation masks, **b)** Examples of IW-TSE segmentation masks after the registration process with misclassified background pixels.

Following the registration and the creation of the IW-TSE MRI segmentation masks, the whole dataset was screened to determine the minimum and maximum bone coordinates. Using those, a rectangular cropping box was created for each IW-TSE MRI. Accordingly, the size of the final IW-TSE MRIs slices was 250 x 320.

### 2.2. Deep learning models and their performance

#### 2.2.1. ResNet-50

The AUC and the PR-AUC values of the highest-performing ResNet-50 models are given in Table 1. The highest AUC (= 0.651, Table 1) and PR-AUC (= 0.379, Table 1) values were achieved when the output probabilities of the ResNet-50 with a batch size of 8 were combined with the patient data (*i.e.*, age, BMI, gender) for the training and testing of the LRC. In this case, the ResNet-50 model was trained on original IW-TSE MRIs with regularization. Next to the AUC and PR-AUC curves, the confusion matrix for the best-performing classifier that combines the output of the ResNet-50 model with patient data is shown in Figure 7. Regardless of the involvement of a ResNet-50 model, the LRC, when trained only with patient' data, achieved an AUC of 0.6399 and a PR-AUC of 0.3643.

### 2.2.2. DenseNet-121

The AUC and the PR-AUC values of the highest-performing DenseNet-121 models are presented in Table 1. The highest AUC values (Table 1) were achieved when the DenseNet-121 model was trained with the original IW-TSE MRIs and the use of the regularization terms, with and without the incorporation of patient data (*i.e.*, age, gender, BMI) in the LRC. When only the DenseNet-121 outputs were counted, the AUC and PR-AUC values were 0.562 and 0.295, respectively. Besides, the combination of the DenseNet-121 model outputs and patient data reached an AUC of 0.656 and a PR-AUC of 0.371 (Table 1). The AUC and PC-AUC curves along with the confusion matrix of the best-performing DenseNet-121 model with the LRC are given in Figure 7.

**Table 1.** The AUC and PR-AUC values for the highest-performing ResNet-50 and DenseNet-121 models for epochs number = 15, batch sizes of 4 and 8, and Adam's optimizer learning rates of 0.001 and 0.01. Models trained on **O:** original IW-TSE MRIs, **O + R:** original IW-TSE MRIs with regularization, **O + R + BW:** original IW-TSE MRIs with regularization and balanced weights, **O + A:** augmented IW-TSE MRIs, **O + A + R:** original and augmented IW-TSE MRIs with regularization.

| Model Variants | Batch Size | Learn. Rate | ResNet-50 AUC | ResNet-50 PR-AUC | ResNet-50 and patient data AUC | ResNet-50 and patient data PR-AUC | DenseNet-121 AUC | DenseNet-121 PR-AUC | DenseNet-121 and patient data AUC | DenseNet-121 and patient data PR-AUC |
|---|---|---|---|---|---|---|---|---|---|---|
| O | 8 | 0.01 | - | - | - | - | 0.627 | 0.371 | 0.641 | 0.365 |
| O + R | 4 | 0.01 | 0.604 | 0.319 | 0.641 | 0.361 | 0.562 | 0.295 | 0.656 | 0.371 |
| O + R | 4 | 0.001 | - | - | - | - | 0.527 | 0.273 | 0.643 | 0.339 |
| O + R | 8 | 0.01 | 0.598 | 0.312 | 0.651 | 0.379 | - | - | - | - |
| O + R + BW | 4 | 0.01 | - | - | - | - | 0.552 | 0.280 | 0.620 | 0.322 |
| O + A | 8 | 0.001 | - | - | - | - | 0.536 | 0.364 | 0.654 | 0.362 |
| O + A + R | 4 | 0.001 | 0.578 | 0.375 | 0.646 | 0.363 | - | - | - | - |
| O + A + R | 8 | 0.01 | 0.585 | 0.391 | 0.646 | 0.367 | - | - | - | - |

### 2.2.3. CVAE

In total, 80 CVAE models were tested by varying the number of epochs (*i.e.,* 25, 50, 100, 200), the size of a batch (*i.e.*, 4, 8), the learning rate (*i.e.*, 0.001, 0.01), the discriminative penalty (*i.e.*, 0.0005, 0.001, 0.01) in the loss function, and the latent feature space (*i.e.*, 100, 500, 1000). CVAE models with the highest AUC values are presented in Table 2. Without the addition of patient data, the best-performing CVAE model reached an AUC of 0.669 and a PR-AUC of 0.346 (Table 2). Together with the patient data, these values slightly increased to 0.670 and 0.359, respectively.

**Table 2.** The AUC and PR-AUC values for the highest-performing CVAE models trained on the balanced IW-TSE MRI dataset with and without regularization terms, with epoch numbers = 100. Learn.: learning, Disc. Pen.: discriminative penalty, Dim.: dimension.

| | | | | Without Kernel Regularization | | | | With Kernel Regularization | | | |
| | | | | CVAE | | CVAE and patients' data | | CVAE | | CVAE and patients' data | |
| Batch Size | Learn. Rate | Disc. Pen. | Latent Space Dim. | AUC | PR-AUC | AUC | PR-AUC | AUC | PR-AUC | AUC | PR-AUC |
|---|---|---|---|---|---|---|---|---|---|---|---|
| 4 | 0.01 | 0.01 | 1000 | 0.669 | 0.346 | 0.670 | 0.359 | - | - | - | - |
| 4 | 0.001 | 0.01 | 1000 | 0.647 | 0.362 | 0.656 | 0.365 | - | - | - | - |
| 4 | 0.001 | 0.01 | 500 | - | - | - | - | 0.601 | 0.416 | 0.614 | 0.410 |
| 4 | 0.001 | 0.001 | 100 | - | - | - | - | 0.565 | 0.290 | 0.629 | 0.329 |
| 4 | 0.01 | 0.001 | 1000 | 0.644 | 0.363 | 0.647 | 0.365 | - | - | - | - |
| 4 | 0.01 | 0.001 | 100 | 0.601 | 0.316 | 0.663 | 0.352 | - | - | - | - |
| 8 | 0.01 | 0.001 | 100 | - | - | - | - | 0.571 | 0.298 | 0.628 | 0.338 |
| 8 | 0.001 | 0.01 | 1000 | - | - | - | - | 0.606 | 0.404 | 0.622 | 0.437 |

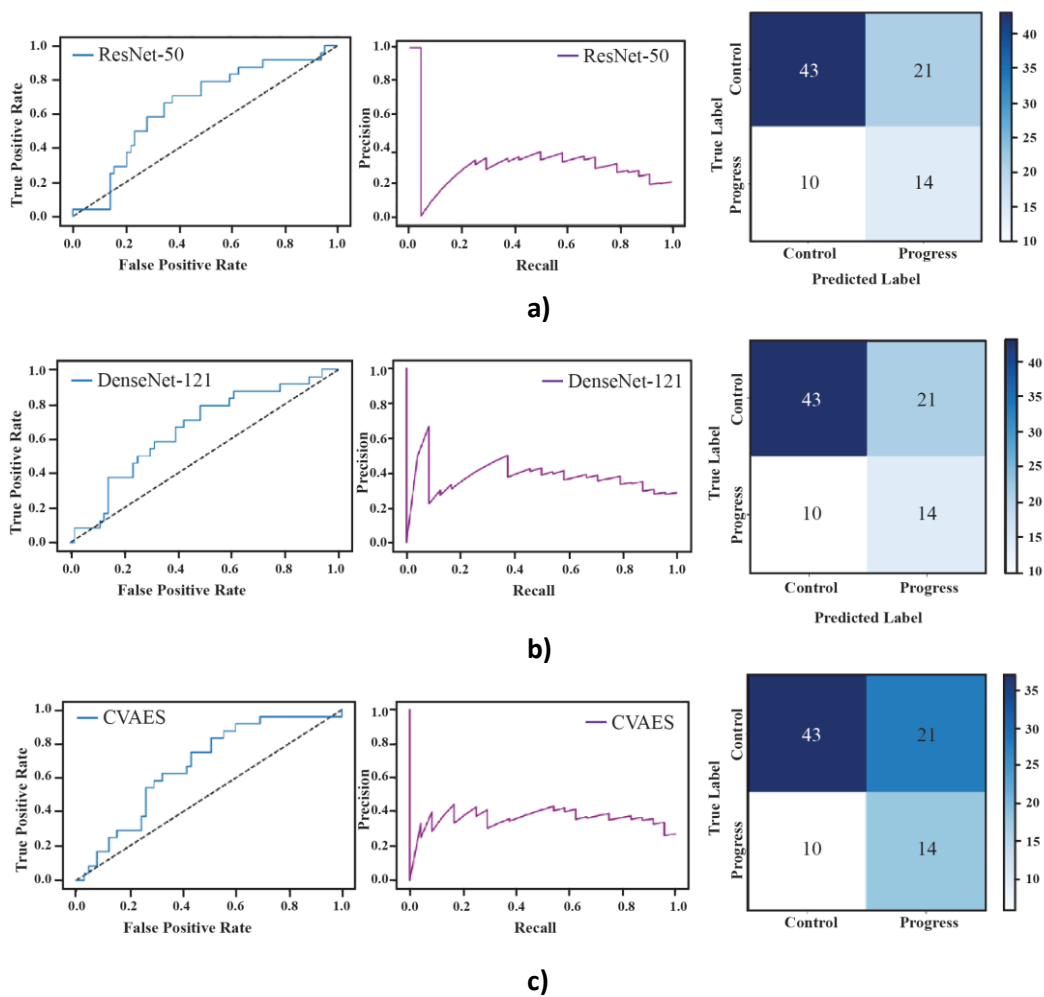

**Figure 7.** The AUC (left column) and PC-AUC (middle column) curves as well as the confusion matrices (right column) for the best performing **a)** ResNet-50 model, **b)** DenseNet-121 model, and **c)** CVAE model.

## 3. Discussion

The goal of this study was to examine the performance of deep-learning models to predict knee OA incidence. With this aim, the DESS and IW-TSE MRIs of subjects who did not have OA progression within the first 24 months from their initial visit (*i.e.*, baseline) and who had it were retrieved from the OAI database and post-processed. The post-process MRIs were used to train, validate and test three different models: ResNet-50, DenseNet-121, and CVAE. To examine the effects of non-image-based features on the prediction of OA incidence, the outputs of the three models were combined with patient data (*i.e.*, age, gender, and BMI) and given as input to an LRC.

The MRI post-processing steps consisted of the segmentation of DESS MRIs, their registrations to IW-TSE MRIs, and the extraction of the area around the knee joint on IW-TSE MRIs. Regarding the segmentation with the U-Net model, the highest mean Dice similarity coefficient of all classes and the mean IoU values were 0.99 and 0.91, respectively, when the model was trained with a batch size of 16. The performance of the U-Net model in segmenting knee structures is in line with that reported by Ambellan *et al.* [18]. For the same batch size, the IoU values, and Dice similarity coefficients for the tibia and the femur were 0.970 and 0.974, and 0.985 and 0.987, respectively. These performance metrics indicate that the sagittal DESS MRIs could be successfully segmented using the U-Net model. Based on the visual evaluation of the registration results, it can be stated that the bone and cartilage labels could be generated successfully. The misclassified pixels could be removed with the inclusion of a simple edge detection algorithm in the workflow.

Concerning the deep-learning algorithms (*i.e.*, ResNet-50, DenseNet-121, CVAE) tested for their performance in the early prediction of OA incidence based on IW-TSE MRIs, the AUC values of all the models were lower than those reported in the literature for X-ray-based studies. This trend did not also change when combining the outputs of the models with patient data. Most of the previous studies developed deep learning algorithms for the detection, diagnosis, and progression of knee OA based on X-rays, which were mostly retrieved from the OAI database. In one study [8], the outputs of a residual-based network trained based on X-rays were combined with clinical data (*i.e.*, age, gender, BMI, and symptomatic assessment of patients). The reported AUC values reached 0.81 for the prognosis of OA progression. The ability of a DenseNet model to predict OA progression based on X-rays was tested by Guan *et al.* [25]. In their study, the best-performing DenseNet model combined with patient data and radiographic risk factors yielded an AUC of 0.86. In another study, Tolpadi *et al.* [26] investigated the ability to predict total knee replacement from MRI scans within 5 years. They used a DenseNet model and incorporated X-rays, 3D IW-TSE MRIs as well as clinical data. The results revealed that the performance of the DenseNet model based on MRIs (*i.e.*, AUC = 0.89) was similar to that based on X-rays (*i.e.*, AUC = 0.85). Overall, the previous studies focused on various tasks (*e.g.*, detection of OA severity) considering X-rays and/or MRIs often reported AUC values greater than 0.8. Several factors might limit the performance of the models developed in this study (*i.e.*, AUC values reached up to 0.670). Previous studies have used X-ray-based outcomes, while MRI and X-ray-based outcome was used in this study. It could be that changes on MRI represent earlier stages of OA than changes on X-ray. The inclusion of various non-image-based features in studies might also explain the difference observed in the performance of the models. Differently from this study, the previous studies did not only include patients' age, gender, and BMI, but they considered several other clinical data as well, *e.g.*, knee injury and surgery history. Comparing the performance of the models in this study to those previous ones based on MRIs, it might be that the types of MRIs used also play a role in the observed difference. This seems to be plausible by referring to the study performed by Schiratti [11]. In this study, the ability to predict the early progression of OA based on joint space narrowing from MRIs was investigated. 2D IW-TSE and 3D DESS MRIs along with subjects' experienced pain were used as input. Their EfficientNet models achieved an AUC of 0.65 and 0.63 when 2D IW-TSE and 3D DESS MRIs were used, respectively. These reported values are close to the ones found in this study.

Regarding the (dis)similarities of the three models developed in this study, ResNet-50 and DenseNet-121 showed similar performance when only the models' outputs were used. This is expected as the architectures of the two models are alike. Comparing CVAE with ResNet-50 and DenseNet-121, it can be stated that the CVAE model was able to distinguish relatively more informative features from the 3D IW-TSE MRIs.

**Conclusion**

The performance of all three deep learning models, *i.e.*, ResNet-50, DenseNet-121, and CVAE, were close to each other to predict OA incidence based on IW-TSE MRIs alone. Although their performance increased with the inclusion of patient data, the latter seems to be more influential in predicting knee OA as compared to IW-TSE MRI-based features.

**Conflict of Interest**

The authors declare that the study was conducted in the absence of any commercial or financial relationships that could be seen as a potential conflict of interest.